\documentclass[prd,
              preprint,
              superscriptaddress,
              preprintnumbers,
              eqsecnum,
              showpacs,
              nofootinbib,
              nobibnotes,
              noeprint,
              nolongbibliography
              ]{revtex4-2}

\usepackage[linkcolor=blue,
            citecolor=blue,
            urlcolor=blue,
            colorlinks=true,
            breaklinks
            ]{hyperref}

\usepackage{booktabs}
\usepackage[italic]{hepnames}
\usepackage{amsfonts}
\usepackage{amsthm}
\usepackage{bm}
\usepackage{slashed}
\usepackage{dsfont}
\usepackage{tikz}
\usepackage[dvipsnames]{xcolor}
\usepackage[shortlabels]{enumitem}
\usepackage{xspace}
\usepackage{siunitx}
\usepackage{bookmark}
\usepackage{bbm}

\def\1eq#1{Eq.\nobreak\thinspace(\ref{#1})}

\def\2eqs#1#2{Eqs.\nobreak\thinspace(\ref{#1}) and\nobreak\thinspace(\ref{#2})}
\def\3eqs#1#2#3{Eqs.\nobreak\thinspace(\ref{#1}),\nobreak\thinspace(\ref{#2}) and\nobreak\thinspace(\ref{#3})}

\def\fig#1{\hyperref[#1]{Fig.\nobreak\thinspace\ref*{#1}}}
\def\figA#1{\hyperref[#1]{Fig.\nobreak\thinspace\ref*{#1}A}}
\def\figB#1{\hyperref[#1]{Fig.\nobreak\thinspace\ref*{#1}B}}
\def\figC#1{\hyperref[#1]{Fig.\nobreak\thinspace\ref*{#1}C}}

\def\tab#1{\hyperref[#1]{Tab.\nobreak\thinspace\ref*{#1}}}

\def\sect#1{\hyperref[#1]{Sec.\nobreak\thinspace\ref*{#1}}}
\def\appref#1{\hyperref[#1]{App.\nobreak\thinspace\ref*{#1}}}

\def\ie{{\it i.e.}, }
\def\eg{{\it e.g.}, }

\newcommand{\be}{\begin{equation}}
\newcommand{\ee}{\end{equation}}

\newcommand{\bea}{\begin{eqnarray}}
\newcommand{\eea}{\end{eqnarray}}

\def\is{S^{-1}}             

\def\g{\Gamma}              

\def\ga{\Gamma_{\!5}}

\def\gv{\Gamma_{\!\s{V}}}

\def\s#1{{\scriptscriptstyle #1}}

\def\Tr{\textrm{Tr}}

\def\srm#1{{\rm{\scriptscriptstyle #1}}}

\begin{document}

\title{\boldmath Light mesons  
in the symmetric-vertex approximation}

\author{M.N.~Ferreira}
\email{narciso.ferreira@ufrgs.br}
\affiliation{\mbox{Instituto de Física, Universidade Federal do Rio Grande do Sul}, Caixa Postal 15051, 91501-970, Porto Alegre, RS, Brazil}

\author{A.S.~Miramontes}
\email{angel.s.miramontes@uv.es}
\affiliation{\mbox{Department of Theoretical Physics and IFIC, University of Valencia and CSIC}, E-46100, Valencia, Spain}

\author{J.M.~Morgado}
\email{jose.m.morgado@uv.es}
\affiliation{\mbox{Department of Theoretical Physics and IFIC, University of Valencia and CSIC}, E-46100, Valencia, Spain}

\author{J.~Papavassiliou}
\email{joannis.papavassiliou@uv.es}
\affiliation{\mbox{Department of Theoretical Physics and IFIC, University of Valencia and CSIC}, E-46100, Valencia, Spain}
\begin{abstract}

We compute the spectrum of light mesons, composed by up, down, and strange quarks, 
using a symmetry-preserving approximation that permits the inclusion of fully-dressed quark-gluon vertices in the key dynamical equations.
This method is characterized by  
the use of the standard symmetric kinematic configuration as a  seed  
in the corresponding Schwinger-Dyson equation,
yielding finally  
the full kinematic dependence of all 
eight form factors composing the 
transversely-projected 
quark-gluon vertex. The extension of 
this approach to the case of distinct nonvanishing current quark masses is 
discussed, and the compatibility
with the fundamental Ward-Takahashi 
identities demonstrated. 
The  
corresponding Bethe-Salpeter kernel is composed by three different diagrammatic structures, 
which may be deduced from the 
attendant quark gap equation by 
applying the standard ``cutting" rules. 
The masses of the light mesons are computed 
by first determining the eigenvalue 
of the Bethe-Salpeter equation 
as a function of 
Euclidean momenta, and then using the Schlessinger extrapolation method to determine the Minkowski momentum for
which this eigenvalue becomes unity.
The resulting meson masses are in good agreement with experimental values, and substantially improve upon predictions from the rainbow-ladder approximation.

\end{abstract}

\maketitle

\newpage 

\section{Introduction}\label{sec:intro}

Recently, a general framework 
was put forth in \cite{Miramontes:2025imd}, which allows for the self-consistent
description of mesons 
within advanced approximation schemes, such as the skeleton or
three-particle irreducible expansions;
for earlier works in this direction, 
see, \eg  \cite{Munczek:1994zz, Matevosyan:2006bk, Fischer:2007ze, Fischer:2008wy, Fischer:2009jm,Chang:2009zb,Sanchis-Alepuz:2015tha,Williams:2014iea,Heupel:2014ina, Sanchis-Alepuz:2014wea, Williams:2015cvx, Sanchis-Alepuz:2015qra, Binosi:2016rxz, Williams:2018adr, Miramontes:2021xgn, Santowsky:2020pwd,Miramontes:2022mex, Gao:2024gdj, Miramontes:2025ofw, Fu:2025hcm, Huber:2025kwy,Miramontes:2025vzb}.
This general approach was 
subsequently streamlined in 
\cite{Ferreira:2025wpu}, 
yielding a tractable set of  
dynamical equations,  fully compatible 
with the constraints imposed 
by the fundamental axial 
Ward-Takahashi identities (WTIs) \cite{Itzykson:1980rh, Miransky:1994vk}, 
emanating from the underlying 
chiral symmetry.

The main technical simplification 
carried out in  
\cite{Ferreira:2025wpu} pertains to   
the Schwinger-Dyson equation (SDE)  
that governs the evolution of 
the quark-gluon vertex, $\g^\mu$; for lattice studies, see \eg \cite{Skullerud:2002sk,Skullerud:2002ge,Skullerud:2003qu,Lin:2005zd,Kizilersu:2006et,Sternbeck:2017ntv,Kizilersu:2021jen,Oliveira:2016muq,Oliveira:2018fkj}. 
In particular, the standard version of this SDE,  used extensively in the 
literature~\cite{Alkofer:2008tt,Williams:2015cvx,Aguilar:2024ciu}, 
is simplified by implementing the 
substitution 
\mbox{$\g^\mu(q,r,-p)= V(q) 
\gamma^\mu$}
\textit{inside} the defining Feynman diagrams, where $q$ denotes the momentum of the gluon.  
The function 
$V(q)$ corresponds to the 
form factor associated with the 
classical tensorial structure, 
evaluated   
in the so-called 
``symmetric'' kinematic configuration, 
defined as \mbox{$q^2=r^2=p^2$}. 
Due to this particular characteristic, we coin this approach as the  
``symmetric-vertex" (SV) approximation. 
Importantly, the evaluation of this  
simplified vertex SDE reproduces rather accurately   
the full momentum-dependence of the corresponding form factors \cite{Gao:2021wun,Ferreira:2025wpu}.

Since the relevant functional equations are 
non-trivially coupled to each other, 
the key practical effect of the
SV approximation is a drastic simplification of the  
Bethe-Salpeter equations (BSEs) that control the formation of mesons within the formalism of \cite{Miramontes:2025imd}; for general works on BSEs, see \cite{Salpeter:1951sz,PhysRev.84.350,Bethe1957,Nakanishi:1969ph,Jain:1993qh,Munczek:1994zz,Maris:1997tm,Alkofer:2002bp, Nicmorus:2008vb,Hilger:2015hka,Hilger:2014nma,Eichmann:2016yit,Mojica:2017tvh,Serna:2017nlr,Wallbott:2019dng,Miramontes:2021exi,Xu:2024fun,Sultan:2024hep,Gutierrez-Guerrero:2024him,Eichmann:2025wgs, Huber:2025cbd,Alkofer:2026vux}.
The final upshot of these considerations 
is the emergence of a symmetry-preserving 
{\it triplet} of dynamical equations, 
namely the SDEs for the quark propagator
(gap equation) 
and quark-gluon vertex, and the meson  
BSEs. In particular, 
in the chiral limit, the results  
obtained from this 
triplet of equations \cite{Ferreira:2025wpu}  satisfy exactly 
the WTI-imposed constraint 
that relates the dominant pion amplitude 
with the quark mass function \cite{Miransky:1994vk}.

In order to further assess the phenomenological potential of this 
approach, 
it is clearly essential to venture 
into the physically relevant 
case of massive mesons. 
Evidently, this task requires 
knowledge of the quark propagator 
and the quark-gluon vertex 
in the complex plane, given that
the mass condition $P^2=-M^2$, with 
$P=(0,0,0, iM)$,  introduces 
complex momenta in the corresponding integrals. However, at present,  
the full complex structure of the QCD 
correlation functions remains largely unexplored; for related works, see, \eg 
\cite{Alkofer:2003jj,Strauss:2012dg,Frederico:2013vga, Dudal:2013yva,dePaula:2016oct, El-Bennich:2016qmb,Tripolt:2018xeo,Dudal:2019gvn,Eichmann:2019dts,Li:2019hyv,Li:2021wol,Fischer:2020xnb,Horak:2021pfr,Horak:2021syv,Horak:2022myj,Eichmann:2021vnj,Horak:2022aza,Huber:2022nzs,Huber:2023uzd,Pawlowski:2024kxc}.

To bypass this difficulty, 
we adopt a method   
used often in the literature in similar circumstances. Specifically, 
the relevant equations are solved for 
an ample number of 
Euclidean values, \ie with $P^2>0$, 
for which the BSE eigenvalue,  
$\lambda(P^2)$, satisfies 
$\lambda(P^2) < 1$. One then  
uses extrapolation techniques,
such as the Pade approximants or the Schlessinger point method (SPM) \cite{Schlessinger:1968vsk, Tripolt:2018xeo, Binosi:2019ecz}, in order 
to determine the Minkowski 
value of $P^2=-M^2$ for which 
$\lambda(P^2)$ becomes unity. The 
error associated with each such mass $M$ is determined by implementing 
the extrapolation using 
distinct samplings of the available 
Euclidean values of $P^2$. This method has been employed in calculations of the meson spectrum \cite{Vujinovic:2014ioa, Eichmann:2019dts, Xu:2022kng, Hagel:2025ngi} and, more recently, extended to investigations of the glueball spectrum \cite{Huber:2020ngt,Huber:2025kwy}.

In the present work we employ the 
procedure described above to 
compute the masses of relatively light mesons, namely mesonic states 
no heavier than about $1.5$ GeV.
Specifically, 
for mesons composed of $u$ and $\bar{d}$ quarks, we compute the masses of \mbox{$\pi^{\pm}$}, \mbox{$\rho(770)$}, \mbox{$b_1(1235)$}, \mbox{$a_1(1260)$}, \mbox{$\pi^{\pm}(1300)$}, and \mbox{$\rho^{\pm}(1450)$}. 
For the strange sector, we calculate the masses of the states \mbox{$K^{\pm}$}, \mbox{$K^*(890)$}, \mbox{$K_{1A}$}, \mbox{$K_{1B}$}, and \mbox{$K^{\pm}(1460)$}.
In general, the computed masses are in good agreement with the experimental values.
In fact, our findings represent a definite improvement over the results obtained 
within the standard rainbow-ladder truncation \cite{Hagel:2025ngi}, where 
the masses of axial-vector mesons and radially excited states tend to deviate considerably from the observed values.

The article is organized as follows. 
In \sect{sec:sva} we briefly review the general theoretical framework of the present approach, and elucidate the salient aspects 
of the SV approximation, paying particular attention to the structure of the BSE-SDE kernel.
In \sect{sec:axialwti} we extend 
certain key demonstrations 
presented in \cite{Ferreira:2025wpu} 
beyond the chiral limit.
In particular, we 
demonstrate that the SDEs 
of the axial-vector and vector vertices,
which share the aforementioned kernel, 
fulfill the correct WTIs in the 
presence of nonvanishing current quark masses.
Then, in \sect{sec:cuts} we show how the 
classic criterion of constructing 
symmetry-preserving BSE kernels \cite{Munczek:1994zz}
applies to the case in hand.
In particular, the SV-derived kernel may 
be also obtained through the appropriate 
functional differentiation 
(``cutting") of the corresponding 
quark self-energy. 
In \sect{sec:vuvs} we discuss the derivation 
of the function $V(q)$ for the up and strange quarks, and present the 
form factors of the corresponding quark-gluon vertices.
In \sect{sec:input} we present the 
results obtained for the meson masses, 
based on the combination of Euclidean analysis and extrapolation techniques described above. 
Then, in \sect{sec:Disc} we summarize our conclusions. 
Finally, in 
\appref{sec:ren} 
we elucidate
on the subtleties associated with the proper renormalization  
in the presence of current quark masses.
 
\section{The symmetric-vertex approximation}\label{sec:sva}

Our considerations commence with  
the axial-vector vertex, $\ga^{a \mu} 
=t^a\ga^{\mu}$,
where $t^a$  
are the generators of the flavour $SU(N_f)$ algebra;
for $N_f=2$, $t^a = \sigma^a/2$, 
while for $N_f=3$, 
$t^a = \lambda^a/2$,
 where $\sigma^a$ 
and $\lambda^a$
are the Pauli and Gell-Mann matrices, respectively. 
Note that  
this vertex is associated with the flavour non-singlet current 
$j^{a\mu}_5(x)=\bar{\psi}(x)t^a\gamma_5 \gamma^\mu{\psi}(x)$, 
 see discussion in Appendix~B of~\cite{Miramontes:2025imd}.
 
In the limit of vanishing current quark masses ($m=0$), the vertex $\ga^{\mu}$
satisfies the well-known WTI \cite{Itzykson:1980rh,Miransky:1994vk}
\be
\label{eq:wtig5}
-P_\mu\ga^\mu(P,p_2,-p_1)=\is(p_1)\gamma_5+\gamma_5\is(p_2)\,,
\ee
where $S^{ab}(p)=i\delta^{ab}S(p)$
represents the 
quark propagator. 
According to the usual decomposition,
$S^{-1}(p)=A(p)\slashed{p}-B(p)$, 
where $A(p)$ and $B(p)$ are the dressings of the (Dirac) vector and scalar structures,  respectively, and 
\mbox{${\mathcal M}(p) = B(p)/ A(p)$} 
is the renormalization-group invariant (RGI) constituent quark mass. 

In order to faithfully capture the 
effects of the underlying chiral symmetry, 
the WTI of \1eq{eq:wtig5} must be preserved when truncations or approximations are implemented. This pivotal requirement, in turn,  
connects inextricably the SDE 
satisfied by $\ga^\mu(P,p_2,-p_1)$ with the gap equation that determines the quark propagator, and constrains the 
form of their most important common ingredient, namely the quark-gluon vertex 
$\g^\mu(q,r,-p)$.

\begin{figure}[t]
    \hspace*{-1cm}
    \includegraphics[scale=0.9]{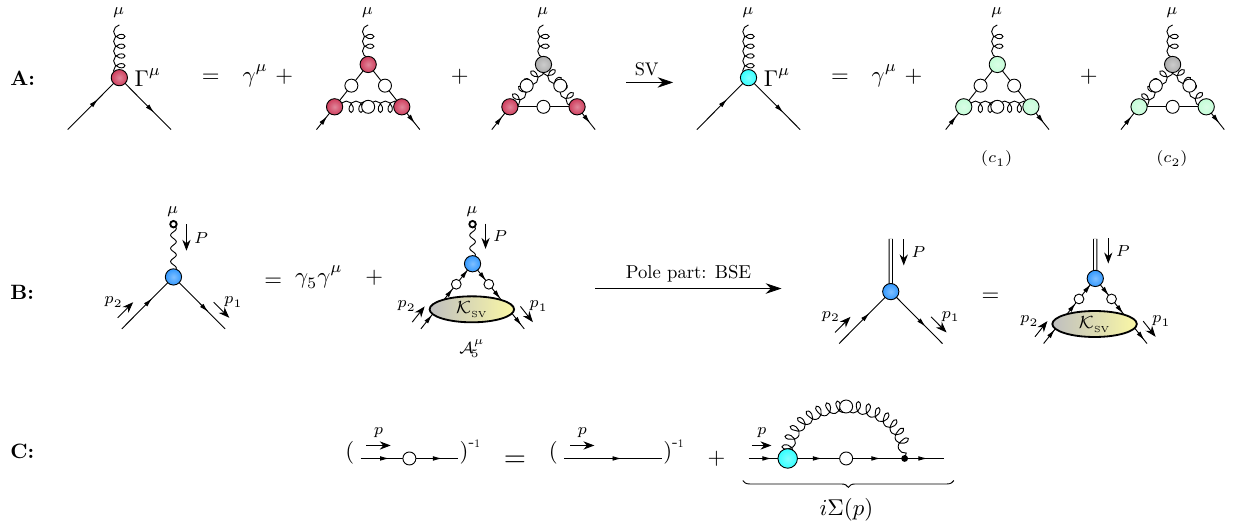}
    \caption{\textbf{\textit{Panel A:}} Diagrammatic representation of the SV approximation at the level of the SDE for the quark-gluon vertex: the r.h.s. arises by implementing into the l.h.s. the substitution indicated in  \1eq{eq:qgsym}.
    \textbf{\textit{Panel B:}} Graphical representation of the axial-vector vertex SDE, within the same approximation, and its connection with the BSE. 
    The kernel $\mathcal{K}_{\!\s{\textrm{SV}}}$ is shown in \fig{fig:bse}. 
    \textbf{\textit{Panel C:}} Pictorial representation of the quark gap equation.}
    \label{fig:sva}
\end{figure}

The key element of the 
SV approximation, put forth in 
\cite{Ferreira:2025wpu},
is the use of a simplified form of the SDE that governs the 
quark-gluon vertex $\g^\mu$. 
In particular, the starting point is the 
version of the SDE for  $\g^\mu$ 
obtained within the three-particle irreducible (3PI) effective action formalism~\cite{Alkofer:2008tt,Blum:2014gna,Williams:2015cvx,Aguilar:2024ciu}, see \figA{fig:sva}.  Equivalently, one may start from the standard  
one-loop dressed SDE, and 
implement the  
skeleton expansion 
to eliminate 
the multiplicative vertex renormalization constant, $Z_1$, 
in favor of the fully-dressed quark-gluon vertex~\cite{Bjorken:1965zz,
Aguilar:2014tka,Ferreira:2024czk,Gao:2024gdj}. 
We emphasize that we do not employ
functional relations derived from the minimization 
of a 3PI effective action
\cite{Cornwall:1973ts,Cornwall:1974vz,Berges:2004pu,Carrington:2010qq,Williams:2015cvx}, 
but, rather, the conventional SDEs, obtained from the functional differentiation of the 
generating functional $Z(J)$~\cite{Itzykson:1980rh,Rivers:1987hi,Alkofer:2008nt,Swanson:2010pw,Huber:2018ned,Huber:2025cbd}.
However, as was demonstrated in~\cite{Miramontes:2025imd}, the direct 
``one-loop dressed" 
expansion of the SDE kernel of 
$\ga^\mu$, together with the 
requirement that the axial WTIs be satisfied exactly, 
lead naturally to the 3PI 
(or skeleton expanded) version of the SDE for $\g^\mu(q,r,-p)$.

The SV approximation amounts to the 
replacement
\be\label{eq:qgsym}
 \g_\mu(q,r,-p) \to V_\mu(q)\,, \qquad \qquad    V_\mu(q):=\gamma_\mu V(q)=\hspace{-4.5cm}\raisebox{-1.25cm}{\includegraphics[scale=1]{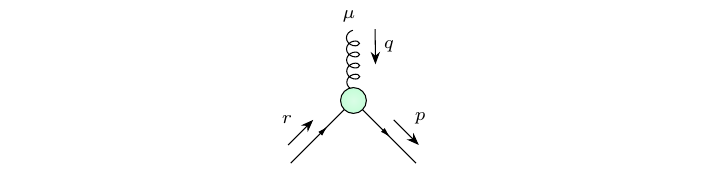}}\hspace{-4.5cm}\,,
\ee
inside the Feynman diagrams
on the r.h.s of the vertex
SDE, as shown in \figA{fig:sva}.

Within this approximation, 
the integral form of the $\g^\mu(q,r,-p)$,  denoted by the cyan vertex, is given by 
\be\label{eq:qgsde}
\g^\mu(q,r,-p) =
\gamma^\mu+c_1^\mu+c_2^\mu\,,
\ee
with
\bea\label{eq:c1c2}
c_1^\mu & = & c_a\int_k V^\beta(k)S(p+k)V^\mu(q)S(r+k)V^\alpha(k)\Delta_{\alpha\beta}(k)\label{eq:c1}\,,\\
\nonumber\\
c_2^\mu & = & c_b\int_k V^\beta(k-q)S(r+k)V^\alpha(k)\Delta_{\rho\beta}(k-q)\Delta_{\alpha\delta}(k)\g^{\mu\delta\rho}(q,-k,k-q)\label{eq:c2}\,,
\eea
where $c_a=-ig^2(C_f-C_A/2)$, $c_b=-ig^2C_A/2$; $C_f$ and $C_A$ are the eigenvalues of the Casimir operator in the fundamental and adjoint representations, respectively [\mbox{$C_f=(N^2-1)/2N$} and \mbox{$C_A=N$} for \mbox{$SU(N)$}]. 
In the 
above formula, 
$\g^{\mu\delta\rho}$ denotes the full three-gluon vertex, 
$\Delta_{\mu\nu}(q)$ stands for the full gluon propagator in the 
Landau gauge, 
\begin{align}
\Delta_{\mu\nu}(q) =  P_{\mu\nu}(q) \Delta(q) 
\,, \qquad\qquad P_{\mu\nu}(q)= g_{\mu\nu} - 
\frac{q_{\mu}q_{\nu}}{q^2} \,,
\label{eq:gluoprop}
\end{align}
and 
we have employed the short-hand notation 
\begin{align}
\int_k :=  \int_{-\infty}^{+\infty}\frac{{\rm d}^4 k}{(2\pi)^4} \,,
\end{align}
where the use of a symmetry-preserving regularization scheme is implicitly assumed.

\begin{figure}[t]
    \hspace*{-1.25cm}
    \includegraphics[scale=1.1]{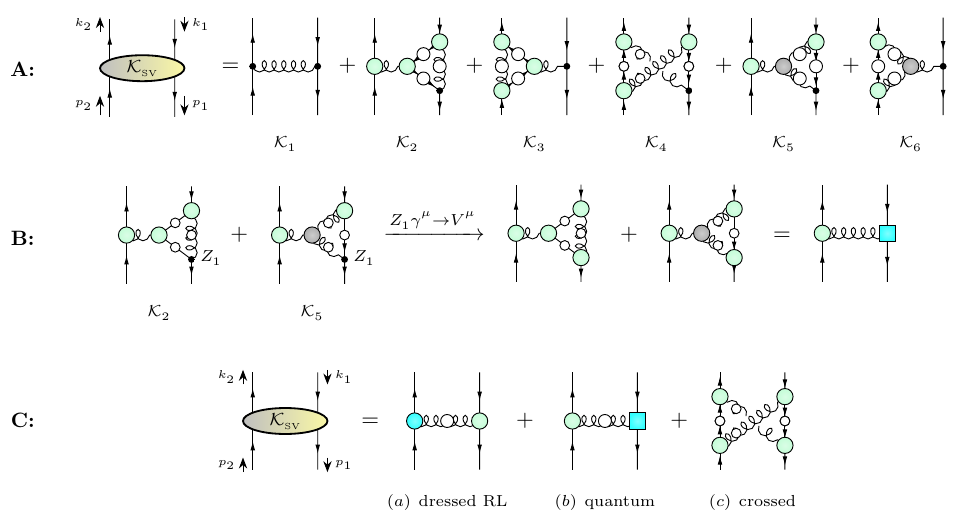}
    \caption{\textbf{\textit{Panel~A:}} The unrenormalized BSE kernel in the SV approximation. \textbf{\textit{Panel~B:}} Schematic representation of the effective 
    renormalization procedure implemented on the kernel \cite{Ferreira:2025wpu}.
    \textbf{\textit{Panel~C:}} The renormalized BSE kernel, where 
    \2eqs{K123}{eq:qgquant} have been used.}
    \label{fig:bse}
\end{figure}

The triplet of symmetry-preserving 
functional equations arising from this 
treatment is shown diagrammatically 
in  \fig{fig:sva}. 
A central component in the ensuing analysis 
is the BSE kernel, denoted by $\mathcal{K}_{\!\s{\textrm{SV}}}$, which essentially defines the 
SDE of the axial-vector vertex, shown on the 
l.h.s. of \figB{fig:sva}. 
The dynamical breaking of the chiral symmetry 
forces $\ga^{\mu}$ to contain a longitudinally-coupled massless pole, associated with the 
attendant Goldstone boson (pion). 
When the pole 
parts of this SDE are singled out, 
one obtains the BSE that governs the formation of the composite pion. As shown on the r.h.s. of \figB{fig:sva}, this homogeneous integral equation involves precisely the kernel $\mathcal{K}_{\!\s{\textrm{SV}}}$.  

Focusing on $\mathcal{K}_{\!\s{\textrm{SV}}}$,
before renormalization
it is given by the sum of the diagrams denoted 
by ${\cal K}_{i}$ in \figA{fig:bse}. 
To see how the cyan vertex arises, 
notice that, by virtue of \1eq{eq:qgsde},  
\be
\label{K123}
 \mathcal{K}_1+\mathcal{K}_3+\mathcal{K}_6=\hspace{-0.25cm}
 \raisebox{-1.5cm}{\includegraphics[scale=1.2]{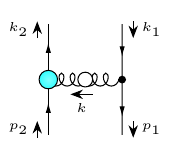}}\,.
\ee
The renormalized BSE kernel $\mathcal{K}_{\!\s{\textrm{SV}}}$
is reached by implementing the 
multiplicative renormalization through the effective procedure described in \cite{Ferreira:2025wpu}  (see also \cite{Fischer:2003rp,Aguilar:2010cn,Aguilar:2018epe}), shown schematically in \figB{fig:bse}. 
In particular, 
the replacement $Z_1\gamma_\mu\to V_\mu$
allows the graphs  ${\cal K}_{2}$ and ${\cal K}_{5}$
to be expressed in terms of the 
 ``quantum'' part, $\g^\mu_{\!\scriptscriptstyle{Q}}(q,r,-p)$, of the 
quark-gluon vertex, defined 
and diagrammatically represented as 
\be\label{eq:qgquant}
 \g^\mu_{\!\s{Q}}(q,r,-p) = c_1^\mu+c_2^\mu
 \,, \qquad \qquad    \g^\mu_{\!\s{Q}}(q,r,-p) := \hspace{-0.75cm}
 \raisebox{-1.5cm}{\includegraphics[scale=1.2]{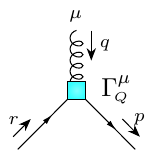}}\,.
\ee

Once the rearrangements described above have been carried out, the kernel $\mathcal{K}_{\!\s{\textrm{SV}}}$ assumes the final form  
shown in \figC{fig:bse}, given by three distinct diagrammatic structures. 
In particular, 
diagram ($a$) is
coined “dressed RL” because it corresponds to the standard RL diagram, but now dressed
with a full quark-gluon vertex. Diagram ($b$) is called “quantum” because it contains the quantum part of this vertex, defined in 
\1eq{eq:qgquant}. Lastly, 
diagram ($c$), is named “crossed” due to its geometry; it contains
only the $V(q)$, and its inclusion is crucial for preserving the WTIs.

We close this section by elaborating on the 
form of the function $V(q)$ in 
\1eq{eq:qgsym}; further details may be found in \sect{sec:vuvs}.
The determination of the 
$V(q)$ is realized through a two-step procedure.
First, the SDE on the l.h.s. of 
\figA{fig:sva} is solved iteratively, maintaining the 
full momentum-dependence of the vertices inside the 
diagrams \cite{Aguilar:2024ciu}. Then, the form factor $\lambda_1(q,r,p)$,
associated with the tree-level (classical) tensor 
$\gamma_{\mu}$, is considered, and 
its slice corresponding to the 
symmetric configuration, $q^2=r^2=p^2$ is singled out,
and identified with $V(q)$. The second step 
is to substitute this $V(q)$ into the
integral expressions for $c_1^\mu$ and $c_2^\mu$ 
in \2eqs{eq:c1}{eq:c2}, and 
multiply by the appropriate projectors 
in order to extract the eight form factors $\lambda_i(q,r,p)$. Note that, in this case, 
the results are not obtained iteratively, but rather 
through simple integrations over the virtual momenta
$d^4k$ in \2eqs{eq:c1}{eq:c2}. 
In this way, 
one obtains the SV approximation to the 
vertex form factors $\lambda_i(q,r,p)$, which display 
full kinematic dependence on the corresponding kinematic variables, \eg 
$r^2$, $p^2$, and the angle $\theta_{rp}$ (Euclidean space). 
As was shown in \cite{Ferreira:2025wpu}, the difference between the 
form factors in the first and second step is relatively small; in that sense, 
the SV approximation reproduces rather 
faithfully the bulk of the quark-gluon vertex.

\section{Beyond the chiral limit}\label{sec:axialwti}

In order to simplify the discussion, 
the considerations presented in \cite{Miramontes:2025imd,Ferreira:2025wpu}  
were restricted to the 
case of vanishing 
current quark masses (chiral limit). 
As was already mentioned there, the generalization of the analysis to the case of 
nonvanishing current quark masses 
is relatively straightforward, and 
may be carried out without 
additional conceptual advances. 
In this section we illustrate this point at the level of two key vertices, 
namely the axial-vector vertex
$\ga^{\mu}$, introduced in the previous section, and the vector vertex 
$\gv^{a\mu}=t^a\gv^\mu$, associated 
with the vector current \mbox{$j_{\s V}^{a\mu}(x)=\bar{\psi}_{f'}(x)t^a\gamma^\mu \psi_f(x)$}. 

These two vertices 
are particularly 
important for the ensuing exploration, 
because they exhibit 
all the meson states considered in this work. 
As we will show,  
in the presence of current quark masses, the 
contraction by $P_\mu$ of 
the SDEs that govern $\ga^{\mu}$ and 
$\gv^\mu$
generates precisely the correct 
WTIs. 
This point is essential for the consistent description of the emerging meson states.
For instance, in the case of 
$\ga^{\mu}$, 
the pseudoscalar poles (\eg $\pi$ and $K$),
and the axial-vector states (\eg $a_1$ and $b_1$) are located in the longitudinal and transverse components of the axial-vector vertex, respectively. Therefore, 
the exact preservation of 
the WTI ensures the proper separation of these two channels. Note that the demonstration is carried out within the 
SV approximation, where the kernel 
entering all relevant SDEs is precisely 
$\mathcal{K}_{\!\s{\textrm{SV}}}$.

The starting point of this analysis are 
the WTIs satisfied by these vertices  
in the presence of current quark masses, namely \cite{Itzykson:1980rh, Miransky:1994vk}
\begin{subequations}\label{eq:AWTI} 
\begin{align}
    -P_\mu \ga^{\mu}(P,p_2,-p_1)&=S^{-1}_f(p_1)\gamma_5+\gamma_5 S^{-1}_{f'}(p_2)+(m_f+m_{f'})\ga(P,p_2,-p_1)\label{eq:awti}\,,\\
P_\mu\gv^\mu(P,p_2,-p_1)&=S^{-1}_{f}(p_1)-S^{-1}_{f'}(p_2)+(m_f-m_{f'})\g(P,p_2,-p_1)\label{eq:vwti}\,, 
\end{align}
\end{subequations}
where
$\ga^{a}=t^a\ga$ is the pseudoscalar vertex, associated with the current
\mbox{$j^a_5=\bar{\psi}_{f'}t^a\gamma_5 \psi_f$}, which 
satisfies the relation \mbox{$\partial_\mu j_5^{a\mu}=-i(m_f+m_{f'})t^a j_5$}; and $\g^a=t^a\g$ is the scalar vertex, defined from the current \mbox{$j^a=\bar{\psi}_{f'}t^a\psi_f$}, which satisfies \mbox{$\partial_\mu j^{a\mu}=i(m_{f'}-m_f)j^a$}.

\begin{figure}[t]
    \centering
    \includegraphics[scale=1.5]{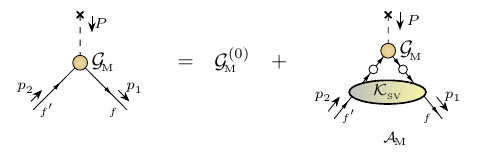}
    \caption{Diagrammatic representation of the generic SDE given in \1eq{eq:sdepseudo}, where 
    \mbox{$\mathcal{G}_{\rm \s{M}} \in \{\g,\ga,\ga^\mu,\gv^\mu \}$} and $\mathcal{G}^{(0)}_{\rm\s{M}} \in
 \{1,\gamma_5, \gamma_5\gamma^\mu, \gamma^\mu\}$. All SDEs have as common ingredient the  kernel $\mathcal{K}_{\s{\textrm{SV}}}$.}
\label{fig:axial}
\end{figure}

In the SV approximation, 
these vertices  
are governed by 
SDEs of the general form shown 
in \fig{fig:axial}, which read   
\be\label{eq:sdepseudo}
\mathcal{G}_{\rm \s{M}}(P,p_2,-p_1)=\mathcal{G}^{(0)}_{\rm \s{M}}-\underbrace{\int_k S_f(k_1)\mathcal{G}_{\rm\s{M}}(P,k_2,-k_1)S_{f'}(k_2)\mathcal{K}_{\s{\textrm{SV}}}(P,k,p_2,-p_1)}_{\mathcal{A}_{\rm\s{M}}(P,p_2,-p_1)}\,,
\ee
with $k_{1,2} := k + p_{1,2}$.
In this compact notation, the index $\rm M$
takes the values ${\rm M = \rm S,P,A,V}$, standing for
``scalar'', ``pseudoscalar'', ``axial-vector'', and ``vector'', so that 
$\mathcal{G}_{\rm \s{M}} \in \{\g,\ga,\ga^\mu,\gv^\mu \}$. The 
 $\mathcal{G}^{(0)}_{\rm\s{M}}$ denotes the corresponding tree-level structures, 
 $\mathcal{G}^{(0)}_{\rm\s{M}} \in
 \{1,\gamma_5, \gamma_5\gamma^\mu, \gamma^\mu\}$. Clearly, the common ingredient of all these SDEs is the kernel 
$\mathcal{K}_{\!\s{\textrm{SV}}}$.

The proof of the WTIs in 
\2eqs{eq:awti}{eq:vwti} proceeds as 
a simple variation of the demonstration 
presented in  \cite{Ferreira:2025wpu} 
for vanishing current quark masses.
Contracting by $P_\mu$ the SDE of the 
axial-vector vertex ($\rm M=A$ in \1eq{eq:sdepseudo}), we get 
\be\label{eq:wtiproof1}
-P_\mu\ga^\mu(P,p_2,-p_1) = (\slashed{p}_1-m_f)\gamma_5+\gamma_5(\slashed{p}_2-m_{f'})+(m_f+m_{f'})\gamma_5-P_\mu\mathcal{A}^{\mu}_5(P,p_2,-p_1)\,,
\ee
where the three first terms on the r.h.s. 
represent the tree-level version of 
\1eq{eq:awti}, while $\mathcal{A}_5^\mu$ 
captures all quantum corrections.

The manipulation of the term 
$P_\mu\mathcal{A}^{\mu}_5(P,p_2,-p_1)$ 
proceeds as in the massless case \cite{Ferreira:2025wpu}:
$P_\mu$ gets contracted with the $\ga^{\mu}(P,k_2,-k_1)$  under the 
integral sign, 
triggering precisely \1eq{eq:awti}. In particular, 
\bea\label{eq:PaQ}
P_\mu\mathcal{A}_5^\mu(P,p_2,-p_1)\!\!&=&\!\!\int_k S_f(k_1)\left[S^{-1}_f(k_1)\gamma_5+\gamma_5 S_{f'}^{-1}(k_2)\right]S_{f'}(k_2)\mathcal{K}_{\s{\textrm{SV}}}(P,k,p_2,-p_1)\nonumber\\
\nonumber\\
&+& 
\!\!(m_f+m_{f'})\!\!\underbrace{\int_{k}S_f(k_1)\ga(P,k_2,-k_1)S_{f'}(k_2)\mathcal{K}_{\s{\textrm{SV}}}(P,k,p_2,-p_1)}_{-\mathcal{A}_5(P,p_2,-p_1)}\,.
\eea

The 
first term on the r.h.s. of \1eq{eq:PaQ} 
generates precisely the appropriate quark self-energies, according to a straightforward procedure described in \cite{Ferreira:2025wpu}, 
which holds even in the presence of 
nonvanishing current quark masses.  In particular, taking into account the 
detailed structure of the kernel $\mathcal{K}_{\s{\textrm{SV}}}$,  
given by the diagrams in \figA{fig:bse}, 
one obtains
\be
\label{eq:PA5mu}
P_\mu\mathcal{A}_5^\mu(P,p_2,-p_1)=i\Sigma_f(p_1)\gamma_5+i\gamma_5\Sigma_{f'}(p_2)-(m_f+m_{f'})\mathcal{A}_5(P,p_2,-p_1)\,.
\ee
Inserting \1eq{eq:PA5mu} into  \1eq{eq:wtiproof1},
and using \1eq{eq:gap}, as well as 
the SDE for $\ga(P,p_2,-p_1)$,  
obtained by setting $\rm M=P$ in  
\1eq{eq:sdepseudo}, 
one recovers precisely the 
WTI of \1eq{eq:awti}.

A completely analogous construction may be followed 
for the vector vertex $\gv^\mu$, which satisfies 
the WTI in \1eq{eq:vwti}.  As before, the starting point is the SDE of the vector vertex, \ie \1eq{eq:sdepseudo} with $\rm M=V$. 
Upon contraction with $P_\mu$, one obtains  
\be\label{eq:vwtitree}
P_\mu\gv^\mu(P,p_2,-p_1)=(\slashed{p}_1-m_f)-(\slashed{p}_2-m_{f'})+(m_f-m_{f'})+P_\mu\mathcal{A}_{\s V}(P,p_2,-p_1)\,.
\ee
where the first terms correspond to the 
tree-level version of \1eq{eq:vwti}, and 
\bea
-P_\mu\mathcal{A}_{\s V}^\mu(P,p_2,-p_1)\!\!&=&\!\!\int_k S_f(k_1)\left[S^{-1}_f(k_1)-S_{f'}^{-1}(k_2)\right]S_{f'}(k_2)\mathcal{K}_{\s{\textrm{SV}}}(P,k,p_2,-p_1)\nonumber\\
\nonumber\\
&+& 
\!\!(m_f-m_{f'})\!\!\underbrace{\int_{k}S_f(k_1)\g(P,k_2,-k_1)S_{f'}(k_2)\mathcal{K}_{\s{\textrm{SV}}}(P,k,p_2,-p_1)}_{-\mathcal{A}_{\!\s{V}}(P,p_2,-p_1)}\,.
\label{eq:PAV}
\eea

Following similar steps as before, 
one can show that 
the first term on the r.h.s. of 
\1eq{eq:PAV} yields the required quark self-energies, such that, 
\be
-P_\mu\mathcal{A}_{\s V}^\mu(P,p_2,-p_1) = 
i\Sigma_f(p_1) - i\Sigma_{f'}(p_2)
+(m_f-m_{f'})\,\mathcal{A}_{\s V}(P,p_2,-p_1)
\,.
\label{eq:PAV2}
\ee
Substituting \1eq{eq:PAV2} into  
\1eq{eq:vwtitree}, using \1eq{eq:gap}, 
and the SDE for $\g(P,p_2,-p_1)$,  
obtained by setting $\rm M=S$ in  
\1eq{eq:sdepseudo}, we obtain the WTI in \1eq{eq:vwti}.

\section{The ``cutting" procedure in the SV approximation}\label{sec:cuts}

As was explained in detail 
in \cite{Ferreira:2025wpu}, 
the symmetry-preserving BSE kernel shown in \figA{fig:bse}
is obtained by an appropriate 
truncation of 
the SDE that governs the axial-vector 
vertex $\ga^\mu(P,p_2,-p_1)$. 
In this section we present the 
derivation of the same kernel  
by employing directly the 
well-known criterion introduced  in \cite{Munczek:1994zz}, namely 
through appropriate functional differentiation of 
the quark self-energy with respect to the quark propagator.

The BSE of a colour-singlet 
vertex, ${\cal G}_{\s{\textrm{M}}}$, which 
may exhibit meson bound states,
has the general form (Euclidean space) \cite{Binosi:2016rxz}
\be
{\cal G}_{\s{\textrm{M}}}(p,P)={\cal G}^{(0)}_{\s{\textrm{M}}} \,+ \,
\int_k S(k_1){\cal G}_{\s M}(k,P)S(k_2)\mathcal{K}(p,k,P)\,,
\label{BSEgen}
\ee
where ${\cal G}^{(0)}_{\rm \s M}$ is the 
tree-level expression of the vertex. Evidently, 
\1eq{BSEgen} is the same integral equation  
written in \1eq{eq:sdepseudo}, but with 
$\mathcal{K}_{\s{\textrm{SV}}} \longleftrightarrow  \mathcal{K}$,
and a set of kinematic variables defined as 
$p_{1,2}=k\pm P/2$ and $k_{1,2}=k\pm P/2$.

The quantity $\mathcal{K}(p,k,P)$ 
represents a kernel that captures all possible interactions between a 
dressed quark and a dressed anti-quark.
$\mathcal{K}(p,k,P)$ is two-particle irreducible; in particular, it does not contain diagrams of a 
quark-antiquark pair annihilating 
into a single gauge boson, nor diagrams that become disconnected by cutting  one quark and one antiquark line.

\begin{figure}[t]
    \hspace*{-1cm}
    \includegraphics[scale=1]{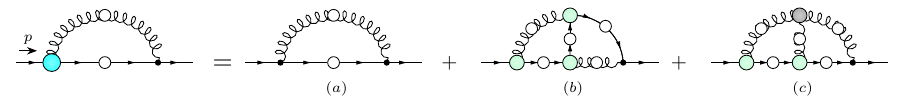}
    \caption{ The diagrammatic expansion of the quark self-energy after the substitution of the cyan vertex by the  
    three diagrams shown in the r.h.s. of \figA{fig:sva}.}
    \label{fig:gapeq}
\end{figure}

The main challenge 
underlying \1eq{BSEgen} 
is to devise  
approximations for 
$\mathcal{K}$
that are compatible with the 
fundamental WTIs of the theory. 
A well-known procedure for constructing such symmetry-preserving approximations for $\mathcal{K}$ 
 was put forth in 
 \cite{Munczek:1994zz}, see also  \cite{Heupel:2014ina,Eichmann:2016yit,Binosi:2016rxz}. 
At its core 
 lies the functional relation 
\be
\label{eq:Kderiv}
\mathcal{K}(p,k,P)=-\frac{\delta\Sigma(p)}{\delta S(k)}\,,
\ee
where $\Sigma(p)$ denotes the 
quark-self-energy,
defined as 
\be\label{eq:gap}
S^{-1}(p)=\slashed{p}-m
-i\Sigma(p)\,, 
\ee
or, from the quark gap equation of \fig{fig:gapeq},
\be
\label{eq:sigma}
\Sigma(p)= - g^2C_f\int_q\gamma^\nu S(q)\g^\mu(q-p,p,-q)\Delta_{\mu\nu}(q-p)\,.
\ee

According to the above criterion, 
an approximation implemented at 
level of the gap equation 
is symmetry-preserving 
as long as the corresponding BSE
kernel satisfies \1eq{eq:Kderiv}. 
Operationally, the functional 
differentiation amounts to the  ``cutting'' of each internal quark line in the diagrams contributing to $\Sigma(p)$.
In particular, the direct cutting rule furnishes the BSE kernel in the 
``diagonal" configuration,   
$\mathcal{K}(p,k,0)$ \,\cite{Carrington:2012ea}. 
The general momentum configuration can be obtained following the procedure described in \cite{Bender:2002as, Binosi:2016rxz}: 
in addition to differentiating, 
the functional derivative adds $P$ to the argument of every
quark line through which it is commuted during the application of the product rule.
Note that, since in this work we restrict ourselves to flavour non-singlet channels, 
the cuts associated with the quark loops   
inside the gluon propagator are disregarded 
\cite{Heupel:2014ina,Binosi:2016rxz}.

Turning to the SV approximation, 
it is relatively straightforward to establish that the key relation of \1eq{eq:Kderiv} is indeed satisfied. 
Specifically, the differentiation 
with respect to $S$ 
of the self-energy containing the cyan vertex, see \fig{fig:gapeq}, gives rise precisely to the 
BSE kernel $\mathcal{K}_{\s{\textrm{SV}}}$, shown in \figA{fig:bse}. 

To see how this occurs, one
employs \1eq{eq:qgsde}
into the r.h.s of 
\1eq{eq:sigma}, thus 
substituting the cyan vertex by the diagrams appearing on the r.h.s. of 
the SDE in \figA{fig:sva}; 
the resulting form of the 
quark self-energy is shown in \fig{fig:gapeq}. 
Then, one carries out 
the aforementioned 
functional differentiation, 
which gives rise to a set of 
``cut" diagrams. In particular,
diagram $(a)$ yields a single cut, 
denoted by ${\cal C}_{\textrm{RL}}$, diagram $(b)$ generates 
three, ${\cal C}_1(b)$, 
${\cal C}_2(b)$, and ${\cal C}_3(b)$, while diagram 
$(c)$ produces two cuts, 
${\cal C}_1(c)$ and 
${\cal C}_2(c)$.

\begin{figure}[t]
    \hspace*{-0.75cm}
    \includegraphics[scale=1]{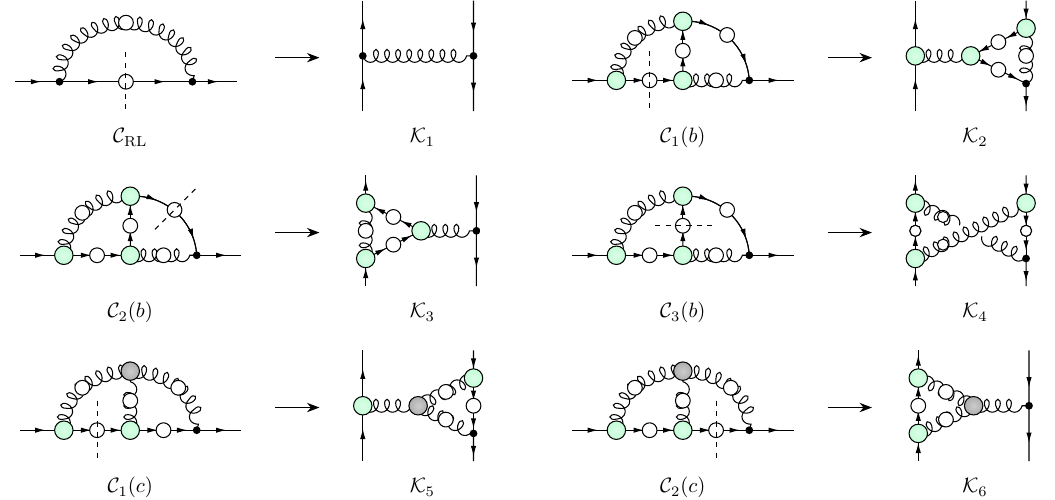}
    \caption{The functional differentiation 
    (``cutting") of the quark self-energy diagrams shown in \fig{fig:gapeq}, leading to the emergence of all the kernel diagrams given in \figA{fig:bse}.}
    \label{fig:cuts}
\end{figure}

The corresponding cuts, 
together with their interpretation as pieces of the unrenormalized BSE kernel, are shown 
in \fig{fig:cuts}. 
Evidently, ${\cal C}_{\textrm{RL}}$ 
corresponds to the standard 
RL case
\be
\mathcal{C}_{\textrm{RL}}\longrightarrow\mathcal{K}_1\,.
\ee
The cuts emerging from diagram $(b)$ of the quark self-energy correspond to the 
terms $\mathcal{K}_i$ that do not contain
a three-gluon vertex, namely 
\begin{align}
    \mathcal{C}_1(b) & \longrightarrow \mathcal{K}_2\,,& \mathcal{C}_2(b)& \longrightarrow  \mathcal{K}_3\,,& \mathcal{C}_3(b)& \longrightarrow\mathcal{K}_4\,.
\end{align}

Finally, the cuts of diagram $(c)$ 
generate the two contributions to the unrenormalized BSE kernel featuring a non-Abelian vertex, namely 
\begin{align}
\mathcal{C}_1(c)&\longrightarrow \mathcal{K}_5\,,& \mathcal{C}_2(c)&\longrightarrow\mathcal{K}_6\,.   
\end{align}

Clearly, the sum of all these six contributions gives rise to the $\mathcal{K}_{\!\s{\textrm{SV}}}$, shown in \figA{fig:bse}. Thus, the BSE kernel 
derived in \cite{Ferreira:2025wpu} 
can be obtained through the cutting procedure of \cite{Munczek:1994zz}.  Therefore, it is both symmetry-preserving 
and universal, in the sense that 
it is common to all mesonic states, regardless of 
their specific quantum numbers \cite{Eichmann:2016yit}.

\section{System of gap equation and quark-gluon vertex}\label{sec:input}

As already mentioned in \sect{sec:sva}, the SV approximation gives rise to a symmetry-preserving triplet of dynamical equations, 
shown in \fig{fig:sva}, 
consisting of the SDE for the 
({\it i}) quark-gluon vertex (panel A), ({\it ii}) the BSE for the mesons (panel B), and ({\it iii}) the quark gap equation (panel C). 
If one neglects possible back-reaction effects of  
the bound states into the structure of 
({\it i}) and ({\it iii}), see, \eg  \cite{Fischer:2007ze,Fischer:2008wy,Mitter:2014wpa, Braun:2014ata, Rennecke:2015eba, Cyrol:2017ewj, Fu:2019hdw, Ihssen:2024miv, Fu:2025hcm}, 
then the BSE in ({\it ii}) may be treated in isolation, 
using the results of the subsystem 
({\it i}) and ({\it iii}) as inputs. 
In this section we consider this subsystem, and determine the inputs that will be employed for the computation of the 
meson masses, presented in the next section. 

\subsection{External inputs}\label{sec:exin}
\begin{figure}[t]
    \centering
    \includegraphics[width=\textwidth, keepaspectratio]{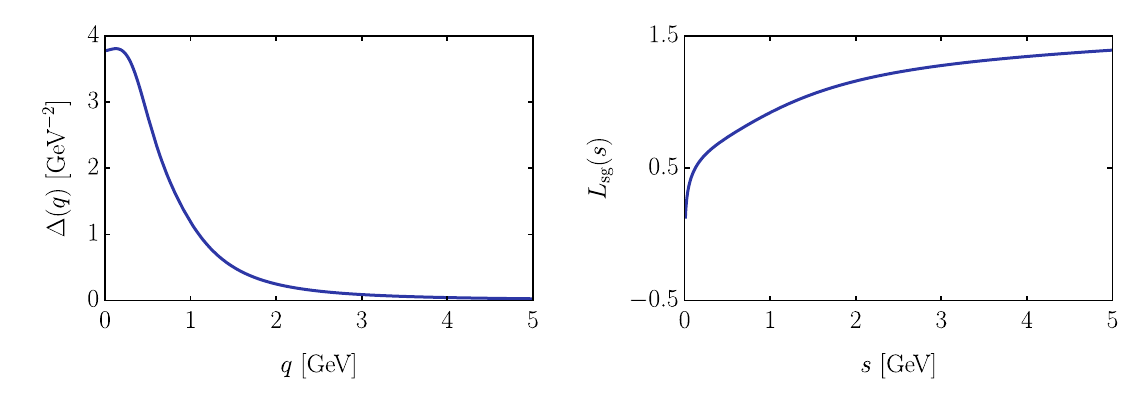}
    \caption{ The external input functions discussed in \sect{sec:exin}.
    \textbf{\textit{Left panel:}} 
  The gluon propagator, corresponding to the 
    lattice data of \cite{Ayala:2012pb}; the functional form of the fit is given in the first line of Eq.~(A1) in  \cite{Aguilar:2023mam}. 
    \textbf{\textit{Right panel:}} 
   The form factor $L_{sg}(s)$
   of the three-gluon vertex, 
corresponding to the lattice data of \cite{Aguilar:2019uob};
the fit is    
   given in the third line of Eq.~(A1) in  \cite{Aguilar:2023mam}.}
    \label{fig:input}
\end{figure}

The treatment of these two equations 
requires the following two external inputs:

({\it a}) The Landau gauge gluon propagator, defined in \1eq{eq:gluoprop}; for its scalar part, $\Delta(q)$, we employ the fit 
given in the first line of Eq.~(A1) in  \cite{Aguilar:2023mam},
shown 
in the left panel of \fig{fig:input}, 
renormalized at $2~\textrm{GeV}$.

({\it b}) The three-gluon vertex, $\g^{\mu\nu\rho}(q,r,p)$, enters the dynamical system through the quark-gluon vertex SDE, \fig{fig:sva}, by means of diagram $c_2^\mu$ in \1eq{eq:qgsde}. In the Landau gauge,  one considers the transversely-projected vertex \mbox{$\overline{\g}^{\mu\nu\rho}(q,r,p)=P^\mu_\alpha(q)P^\nu_\beta(r)P^\rho_\gamma(p)\g^{\alpha\beta\gamma}(q,r,p)$}.
Due to the planar degeneracy of this vertex
\cite{Blum:2014gna,Eichmann:2014xya,Ferreira:2023fva,Aguilar:2023qqd,Pinto-Gomez:2022brg,Pinto-Gomez:2024mrk},  
the tree-level structure 
\begin{align}
\overline{\g}^{\mu\nu\rho}(q,r,-p)&=
[g^{\nu\rho}(r-p)^\mu+g^{\mu\rho}(p-q)^\nu+g^{\mu\nu}(q-r)^\rho] L_{sg}(s)
\,,
\end{align}
with $s^2=\frac{1}{2}(q^2+r^2+p^2)$, 
is an excellent approximation.  
The form factor $L_{sg}(s)$  
corresponds to the fit given in the third line of Eq.~(A1) in \cite{Aguilar:2023mam}, and is 
 shown in the right panel of \fig{fig:input}.

\subsection{The functions \texorpdfstring{$V_{\!u}(q)$}{Vu(q)} and \texorpdfstring{$V_{\!s}(q)$}{Vs(q)}}\label{sec:vuvs}

Given that we will consider 
mesonic states containing the 
up, the down, and the strange quarks, the function 
$V(q)$ must be determined separately for each 
case, $V_{\!u}(q)$,  $V_{\!d}(q)$, and $V_{\!s}(q)$. 
To that end, the 
gap equation 
and the SDE of the 
quark-gluon vertex on the l.h.s. of 
\figA{fig:sva}
must be solved as a coupled system, once for an up and once for a strange quark.
Evidently, the distinction 
between quark flavours 
is encoded in the 
value of the current quark mass, 
$m_u$ or $m_s$, 
used in the gap equation.
The $d$ quark case does not need to be considered separately, as the limit of exact isospin symmetry is assumed, 
\ie perfect degeneracy between up and down quarks, $m_u = m_d$, and $V_u(q)=V_d(q)$.

In order to carry out the numerical 
treatment of these equations, one must 
employ standard conversion rules to 
pass them from Minkowski to Euclidean space, 
see \eg Sec.~IV~B and App.~A in \cite{Aguilar:2024ciu}). Note also that,
due to the presence of different 
current quark masses, 
the {\it renormalization} of these equations 
requires particular care, in order to 
preserve the compatibility with the fundamental WTIs. The details of the renormalization procedure adopted in this 
work are presented in \appref{sec:ren}.

\begin{figure}
    \centering
    \includegraphics[width=\textwidth, keepaspectratio]{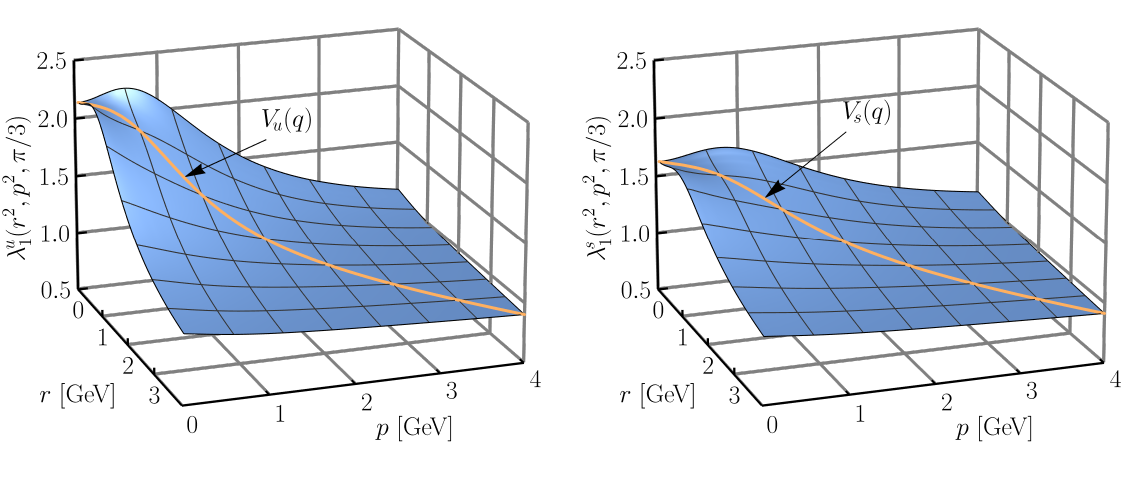}
    \caption{The classical form factors
    $\lambda_1^u(r^2,p^2,\pi/3)$ 
(left panel) and   $\lambda_1^s(r^2,p^2,\pi/3)$  (right panel) of the quark-gluon vertex for an up and a strange quark, respectively; the results are obtained from 
the solution of the SDE on the l.h.s. of 
\figA{fig:sva} (red vertex). The diagonal slices ($r^2=p^2$) are identified with $V_u(q)$ and $V_s(q)$, 
respectively.}
    \label{fig:Vfdet}
\end{figure}

As explained 
at the end of \sect{sec:sva}, 
for the determination of the $V(q)$ we only need to consider the resulting 3-D plots for 
the classical form factors 
$\lambda_1^{\!u}(r^2,p^2,\theta_{rp})$ and $\lambda_1^{\!s}(r^2,p^2,\theta_{rp})$, 
shown in \fig{fig:Vfdet}, 
and identify the slice that 
corresponds to the symmetric 
configuration, $q^2=p^2=r^2$,  
$\theta_{rp} = \pi/3$, \ie  
\begin{align}
    V_{\!u}(q)&:=\lambda_1^{\!u}(q^2,q^2,\pi/3)\,,& V_{\!s}(q)&:=\lambda_1^{\!s}(q^2,q^2,\pi/3)\,.
\end{align}

\begin{table}[!t]
    \centering
    \setlength{\tabcolsep}{10pt} 
    \resizebox{\textwidth}{!}{%
    \begin{tabular}{lcccccccc}
    \toprule\toprule
        &
        $d$ &
        $\kappa$ &
        $b^2_0$ &
        $b^2_1~[\textrm{GeV}^2]$ &
        $b^2_2~[\textrm{GeV}^2]$ &
        $b^2_3~[\textrm{GeV}^2]$ &
        $e^2_0~[\textrm{GeV}^2]$ &
        $e^2_1~[\textrm{GeV}^2]$ \\
        \midrule
        Light \hfill $(f=u)$ & $1.154$  & $1.332$ & $0.0146$ & $126.113$ & $217.268$ & $4.766$ & $3.916$ & $2.079$ \\\hline
        Strange \hfil $(f=s)$ & $1.154$  & $1.103$ & $0.0182$ & $50.581$ & $999.996$ & $3.274$ & $3.916$ & $2.079$ \\
    \bottomrule\bottomrule
    \end{tabular}
    }
    \caption{The values of the parameters 
    used in the fit of 
    \2eqs{eq:V_fit}{fitpar} for 
$V_{\!u}(q)$ and $V_{\!s}(q)$.}
    \label{tab:coefsV}
\end{table}

Both curves may be fitted very accurately 
by a function of the form 
\be
    V(q)=\frac{d}{U(q)^\frac{9}{4\beta_0}+R(q)}\,,
    \label{eq:V_fit}
\ee
with $\beta_0=11-2N_f/3$, and 
\begin{align}
U(q)&=1+\kappa\log\left(\frac{q^2+\eta_f(q)}{\mu^2+\eta(q)}\right)\,, &R(q)&=\frac{b_0^2+q^2/b_1^2}{1+q^2/b_2^2+\left(q^2/b_3^2\right)^2}\,,& \eta_f(q)&=\frac{e_0^2}{1+q^2/e_1^2}\,,
\label{fitpar}
\end{align}
with the optimal values of the fitting 
parameters collected in \tab{tab:coefsV}.

\subsection{Full quark-gluon vertex in the SV approximation}\label{app:qgffs}

\begin{figure}[t]
    \centering
    \includegraphics[scale=0.8]{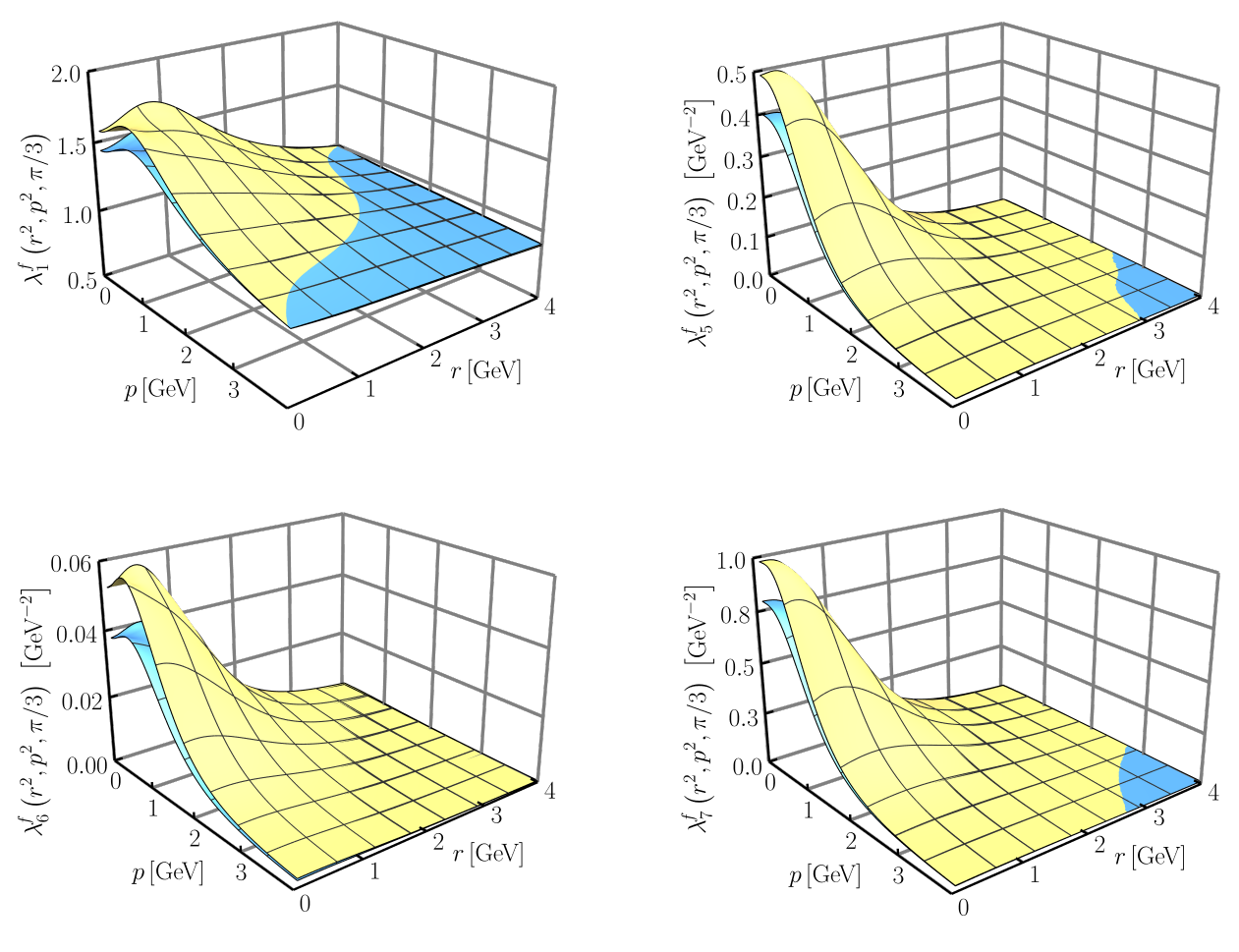}
    \caption{The chirally symmetric form factors $\lambda_i^{f}$ ($i=1,5,6,7$) of the quark-gluon vertex, within the SV approximation. 
    Yellow surfaces correspond to the case $f=u$ (up quark), while blue surfaces to the case  $f=s$ (strange quark).}
    \label{fig:LambdaOddComparison}
\end{figure}

The last step for acquiring all necessary 
ingredients for the ensuing treatment of the
meson BSE is to determine 
the quark-gluon vertex  
using the equation on the r.h.s. of \figA{fig:sva}, 
or, equivalently, by substituting 
into the integral expressions of 
\2eqs{eq:c1c2}{eq:c2}
the $V_{\!u}(q)$ and $V_{\!s}(q)$
obtained above.

We remind the reader that 
the transition from the red to the cyan quark-gluon vertex is required in order to 
simplify the form of the BSE kernel, 
in a symmetry-preserving way. 
Specifically, as explained in 
\cite{Miramontes:2025imd,Ferreira:2025wpu},
the red quark-gluon 
vertex is compatible with the 
BSE kernel that is composed by additional diagrams, containing the so-called ``gluon-axial-vector" vertex, $G_5^{\mu\nu}$. 
The omission of these diagrams, which leads 
eventually to the kernel $\mathcal{K}_{\!\s{\textrm{SV}}}$, 
is compatible with the WTIs 
provided that the cyan quark-gluon vertex 
is used instead of the red one.  

For the actual computation we 
turn to the transversely-projected quark-gluon vertex, 
\be
\overline{\Gamma}_\mu(q,r,-p)= P_{\mu\nu}(q)\Gamma^{\nu}(q,r,-p)\,,
\ee
\begin{figure}[t]
    \centering
    \includegraphics[scale=0.8]{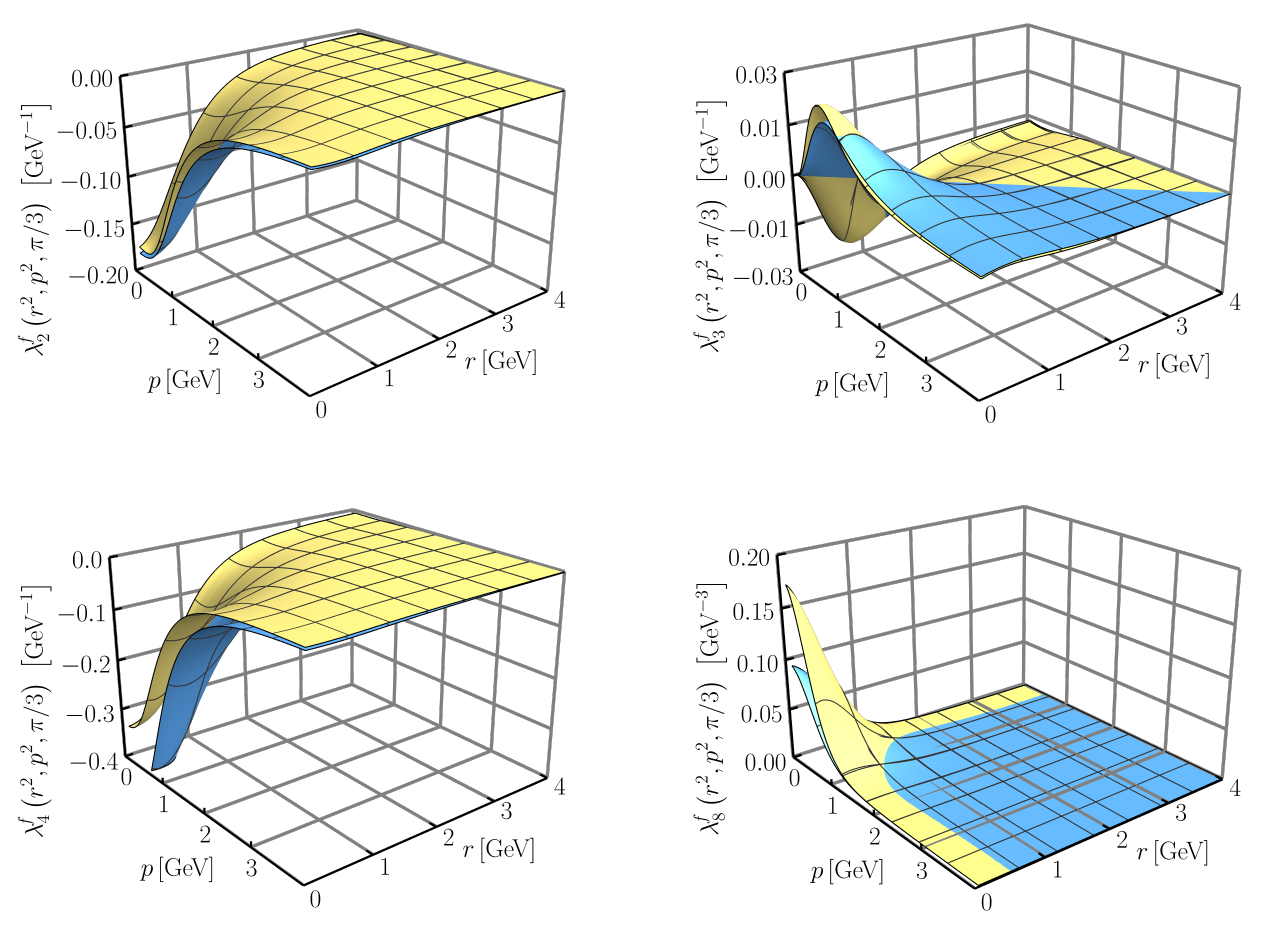}
    \caption{The chiral symmetry breaking form factors $\lambda^f_i$ ($i=2,3,4,8$) of the quark-gluon vertex in the SV approximation. As before,  
    yellow surfaces indicate the case $f=u$ (up quark), while blue surfaces the case  $f=s$ (strange quark).}
    \label{fig:LambdaEvenComparison}
\end{figure}
which can be spanned in a basis of eight independent tensors, $\bar\tau_{i}$, 
\begin{align} 
\label{decomp}
    \overline{\Gamma}^\mu(q,r,-p)=\sum_{i=1}^{8}\lambda_i(q,r,-p)\bar\tau_{i}^\mu(r,-p) \,, 
    \qquad\qquad  
 \bar\tau_{i}^\mu(r,-p)=  P^\mu_{\nu}(q)\tau_{i}^\nu(r,-p) \,,
\end{align}
and the $\lambda_i$ are the attendant form factors. The basis chosen is given by 
~\cite{Mitter:2014wpa, Cyrol:2017ewj, Gao:2021wun, Ihssen:2024miv,Aguilar:2024ciu} 
\begin{align}
\label{Taus}
    &\tau^\nu_{1}(r,-p) =\gamma^\nu\,, \quad &&\tau^\nu_{2}(r,-p) =  (p+r)^\nu\,, \nonumber \\
    &\tau^\nu_{3}(r,-p) = (\slashed{p}+\slashed{r})\gamma^\nu\,, \quad &&\tau^\nu_{4}(r,-p) = (\slashed{p}-\slashed{r})\gamma^\nu\,,\nonumber\\
    &\tau^\nu_{5}(r,-p) =  (\slashed{p}-\slashed{r})(p+r)^\nu\,, \quad &&\tau^\nu_{6}(r,-p) =(\slashed{p}+\slashed{r})(p+r)^\nu\,,\nonumber\\
    &\tau^\nu_{7}(r,-p) = -\frac{1}{2}[\slashed{p},\slashed{r}]\gamma^\nu\,, \quad &&\tau^\nu_{8}(r,-p) = -\frac{1}{2}[\slashed{p},\slashed{r}](p+r)^\nu \,.
\end{align}

We then substitute the $V_{\!u}(q)$ and $V_{\!s}(q)$
into \2eqs{eq:c1c2}{eq:c2}, and employ the projectors given by 
Eq.\,(3.9) of \cite{Aguilar:2024ciu}
to extract the individual form factors 
$\lambda_i^u$ and $\lambda_i^s$. 
The results for the chirally symmetric form factors $\lambda_{1,5,6,7}$ 
are shown in \fig{fig:LambdaOddComparison}, 
while the chiral symmetry breaking 
$\lambda_{2,3,4,8}$ are displayed 
in \fig{fig:LambdaEvenComparison}.

Finally, the Dirac components $A(p)$ and $B(p)$ of the quark propagators,  
together with the RGI quark mass ${\cal M}(p)$, for both the up and strange quarks, 
are presented in \fig{fig:quark_AB} 
of~\appref{sec:ren}. 

\section{Numerical results for meson masses}\label{sec:res}

In this section we present the meson spectrum obtained from the BSE 
constructed in the SV approximation. 
In particular, the 
three main diagrams that compose the 
BSE kernel, $\mathcal{K}_{\!\s{\textrm{SV}}}$, shown in  
\figC{fig:bse}, are put together using 
the ingredients discussed previously.

The BSE displayed in \figC{fig:bse} describes the formation of meson bound states for the channels considered in this work, namely pseudoscalar $(J^{PC} = 0^{-+})$, vector $(J^{PC} = 1^{--})$, and axial-vector mesons $(J^{PC} = 1^{++},1^{+-} )$. What distinguishes the different states is the Dirac structure of the corresponding BS amplitude, which is decomposed into a basis of covariants compatible with the quantum numbers of each state. In the present work, we employ the appropriate Dirac basis for each channel, as given explicitly in Tables I and II of \cite{Fischer:2008wy}. A collection of bases for different spin states may also be found in \cite{Hagel:2025ngi}.

The interaction kernel $\mathcal{K}_{\!\s{\textrm{SV}}}$, described in the previous sections, enters in the BSE for each of the channels considered here. The different physical content of each state is encoded in the specific basis for the BS amplitude, and in the corresponding projection onto the BSE eigenvalue problem described below.

The remaining inputs used in the BSE are the gluon propagator, the quark-gluon vertex, and the functions $V_{\!u}(q)$ and $V_{\!s}(q)$; details about all these quantities are presented in \sect{sec:input}.
\begin{figure}[t]
    \hspace*{-0.5cm}
    \includegraphics[scale=0.65]{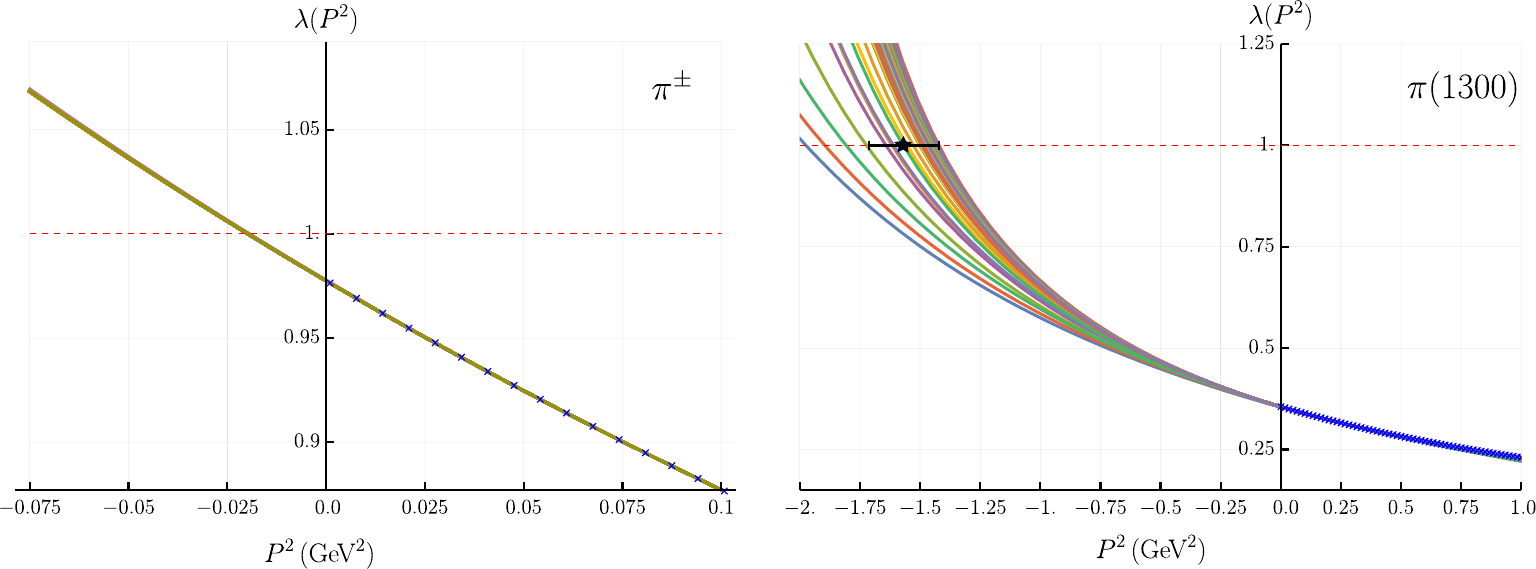}
    \caption{BSE eigenvalue $\lambda(P^2)$ as a function of the total momentum $P^2$, for the $\pi^{\pm}$ (left) and $\pi(1300)$ (right). The dots correspond to  Euclidean data points, the band represents the ensemble of SPM extrapolations, and the horizontal dashed line marks the bound state condition $\lambda = 1$. The horizontal black bar in the right panel indicates the standard deviation of the extracted bound state mass. }
    \label{fig:Eigenvalues}
\end{figure}

In order to determine the meson masses, we use the SPM 
\cite{Schlessinger:1968vsk, Tripolt:2018xeo, Binosi:2019ecz}, in order 
to extrapolate the BS eigenvalue curve $\lambda(P^2)$ from the Euclidean region to the on-shell domain. Specifically, we solve the BSE for a set of $N$ discrete values of the momentum $P$, namely $\{P_1,P_2, ...,P_N \}$,
and obtain the corresponding eigenvalues $\lambda_i(P^2)$. A bound state of mass $M$ is then identified by the condition $\lambda(P^2)=1$ at the pole location $P^2= - M^2$.

To estimate this crossing point we employ a resampling procedure. From the full set of $N$ calculated points in the Euclidean region, we choose a random subset of size $n<N$ and construct an SPM approximant for $\lambda(P^2)$. We then determine the value of $P^2$ where the reconstructed curve fulfills the condition $\lambda(P^2) = 1$. We repeat this process $m$ times, each time using an independently resampled subset, producing an ensemble of mass estimates 
$\{M_1,M_2, ...,M_m \}$. Then, the final result is identified  with the mean value, while the standard deviation 
$\sigma$ is used 
as an estimate of the uncertainty associated with the SPM extrapolation,   
\begin{align}
M&=\frac{1}{m}\sum_{i=1}^m M_i\,,& \sigma&=\sqrt{\frac{1}{m-1}\sum_{i=1}^m(M_i-M)^2}\,.
\end{align}

\begin{figure}[t]
    \includegraphics[scale=0.85]{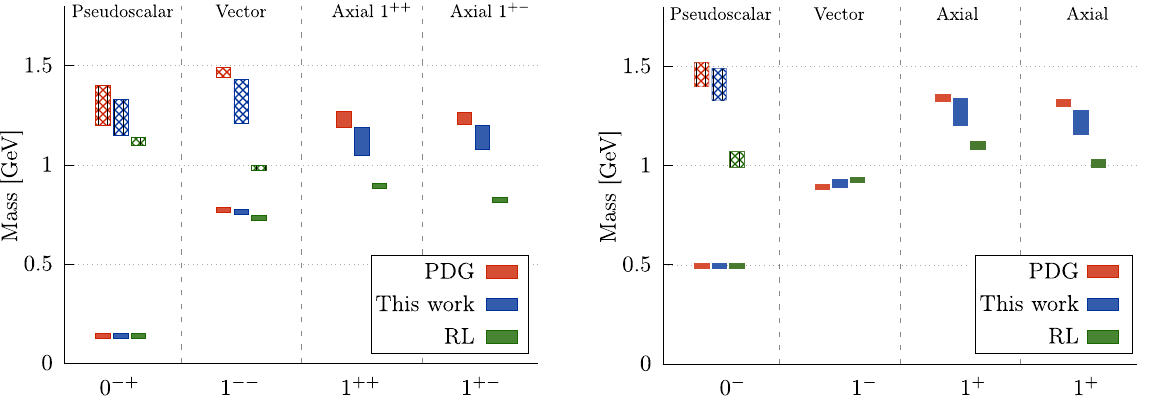}
    \caption{Meson mass spectrum for the $u\bar{d}$ (left) and $u\bar{s}$ (right)  sectors, comparing results from this work, RL~\cite{Hagel:2025ngi},  and PDG values~\cite{ParticleDataGroup:2024cfk}. Solid (hatched) boxes denote ground (excited) states, with box height indicating the mass uncertainty. In the axial $1^+$ kaon channel, the PDG entries correspond to the unmixed $K_{1A}$/$K_{1B}$ basis states extracted from the physical $K_1(1270)$/$K_1(1400)$ masses.}
    \label{fig:spectrum}
\end{figure}

The method described above is employed for the light meson sector, including pseudoscalars, vector and axial-vector states. In \fig{fig:Eigenvalues} we show representative examples of the extrapolation for the pion $\pi$ and the 
radial excitation $\pi(1300)$. Each panel displays the Euclidean data points, computed directly from the BSE, together with the band of SPM reconstructions generated by the resampling procedure. As $P^2$ moves from the Euclidean region towards negative values, $\lambda(P^2)$ grows reaching unity at $P^2 = -M^2$. The width of the band provides a visual representation of the extrapolation uncertainty. 
In particular, in the case of the ground state pion, the curves are well constrained across the range of momentum. 
Instead, for the radially excited channel, the bound state mass is located at a larger value of $P^2$; 
therefore, the extrapolation needs to be implemented over a greater distance from the Euclidean data points. 

A global view of the computed spectrum is presented in \fig{fig:spectrum}, with the corresponding masses collected in detail in \tab{tab:meson_masses}. In both cases, the results are compared with both the RL truncation and with the experimental values reported by the PDG \cite{ParticleDataGroup:2024cfk}. The current quark masses are fixed requiring that the computed $\pi$ and $K$ reproduce their experimental values; so, these two states serve as benchmarks. The remaining masses are predictions of the SV approximation, assisted by the SPM extrapolation.

The overall pattern shows that the SV approximation provides a systematically improved description across the different $J^{PC}$ channels with respect to the RL counterpart. In particular, the degeneracy between the $a_1$ and $b_1$ masses predicted within the present framework is consistent with experimental determinations. 
A further improvement concerns the radial excitations: within the RL truncation, these states are significantly underestimated, and their mass hierarchy is not correctly reproduced with respect to the experimental ordering. The central values obtained with the current approximation suggest a correction of both issues, yielding masses closer to the experimental data, and a more consistent ordering 
among the radial excitations.

It should be noted that, in the strange axial $1^+$ sector, C-parity is not a good quantum number for kaon states, and SU(3) flavour symmetry breaking induces mixing between the $K_{1A}$ and $K_{1B}$ basis states, giving rise to the physical mass states $K_1(1270)$ and $K_1(1400)$~\cite{Suzuki:1993yc,Burakovsky:1997dd,Cheng:2011pb}. 
Our BSE solutions correspond to the unmixed states $K_{1A}$ and $K_{1B}$, as do the RL results. To allow a direct comparison, we extract the unmixed basis state masses from the physical PDG entries, using the mixing angle $\theta = (40\pm 5)^\circ$~\cite{Liu:2024lph}.

\begin{table*}[t]
\centering
\renewcommand{\arraystretch}{0.8}
\resizebox{\textwidth}{!}{%
\begin{tabular}{l ccccccccccc}
\hline\hline
 & $\pi$ & $\rho$ & $a_1$ & $b_1$ & $\pi(1300)$ & $\rho(1450)$
 & $K$ & $K^*$ & $K_{1A}$ & $K_{1B}$ & $K(1460)$ \\
\hline
This work
  & $0.139(1)$ & $0.763(5)$ & $1.12(7)$ & $1.14(6)$ & $1.24(7)$  & $1.32(11)$
  & $0.495(1)$ & $0.91(2)$  & $1.27(7)$ & $1.20(6)$ & $1.41(8)$ \\
RL~\cite{Hagel:2025ngi}
  & $0.139$   & $0.732$   & $0.894$  & $0.822$  & $1.12(2)$   & $0.986$
  & $0.495$   & $0.930$   & $1.10(2)$  & $1.01(2)$   & $1.03(4)$  \\
Exp.
  & $0.139(1)$ & $0.775(1)$ & $1.23(4)$ & $1.230(3)$ & $1.30(10)$ & $1.46(3)$
  & $0.494(1)$ & $0.892(1)$ & $1.343(17)$ & $1.317(17)$ & $1.46(6)$ \\
\hline\hline
\end{tabular}}
\caption{Meson masses (in GeV) compared to RL results from \cite{Hagel:2025ngi},  and
the experimental values from the PDG \cite{ParticleDataGroup:2024cfk}. 
\label{tab:meson_masses}}
\end{table*}

\section{Discussion and conclusions}\label{sec:Disc}

In this work we 
have presented the first comprehensive
study of the meson spectrum using 
the general symmetry-preserving 
truncation introduced in \cite{Ferreira:2025wpu},
and appropriately extended here to 
include the physically relevant case 
of nonvanishing quark masses. 
The main characteristic of this approach  is the use of a fully-dressed quark-gluon vertex, with all its form factors participating actively in the 
formation of the bound states described by the corresponding BSE.

In order to bypass the need to 
extend in the complex plane the Green's function that 
enter in the BSE, we have used the common method \cite{Vujinovic:2014ioa, Eichmann:2019dts, Xu:2022kng, Hagel:2025ngi, Huber:2020ngt,Huber:2025kwy}
of solving the 
corresponding dynamical equations for 
a wide sample of Euclidean $P^2 >0$, 
and then using standard extrapolation techniques to deduce the masses, $P^2 = -M^2$. 
In particular, we 
have focused our attention to the case 
of light quarks, composed by up, down, and strange quarks. We have resorted to 
the SPM \cite{Schlessinger:1968vsk, Tripolt:2018xeo, Binosi:2019ecz}
in order to 
estimate the masses from the solution  of the main system of equations (see 
\fig{fig:sva}) 
in the Euclidean space. 
Our findings are in good agreement 
with the experimental values, and 
represent an overall improvement over 
the results obtained in the recent 
comprehensive RL study of \cite{Hagel:2025ngi}. 

Even though the method employed here is in principle 
applicable to any mesonic state, 
the limitation to the light ones is essentially 
imposed by the use of the SPM, and the error associated with it. 
In particular, as the meson masses 
become heavier,  
the required range of the extrapolation 
increases, and so does the error 
assigned to the final prediction; 
compare, \eg the left and right panels of \fig{fig:Eigenvalues}.

Evidently, other sources of error exist in our analysis, such as the truncation of the corresponding 
functional equations, and the errors associated with the external inputs (gluon propagator and three-gluon form factors, see \sect{sec:exin}). 
These systematic errors are partly absorbed into the values of the coupling and the current quark masses, which were tuned to reproduce the lightest states, as explained in App.~\ref{sec:ren}. However, they should contribute to the error budget of the heavier mesons; nevertheless, we expect the rather large extrapolation errors to be the main source of error in these cases. Finally, there are other numerical errors, such as those incurred by using numerical quadratures and interpolations. These were checked to be small (order 10$^{-3}$) for the Euclidean data, and are naturally included in the SPM extrapolation error, since the small numerical noise in the data is part of the reason why the various SPM approximants (see, \eg \fig{fig:Eigenvalues}) constructed from different samples differ from one another. 

It is important to emphasize that the only adjustable parameters in this entire analysis are those present in the QCD Lagrangian, namely, the values of the strong coupling, $\alpha_s(\mu)$, and the current quark masses, $m_u(\mu)$ and $m_s(\mu)$. While these values were adjusted to reproduce the experimental values of the $\pi$ and $K$ mesons, all other nine calculated meson masses emerge as genuine predictions of the framework. Note, in particular, that the 
curves employed for $V_u(q)$ and $V_s(q)$ are precisely as obtained from the direct solution of the vertex SDE,  
see \fig{fig:Vfdet}, without any fine-tuning or further adjustments. In that sense, 
the agreement seen in \fig{fig:spectrum} and \tab{tab:meson_masses} between our predictions and the experimental values is 
rather nontrivial; it results from the incorporation of the full tensor structure of the nonperturbative quark-gluon vertex, which provides the necessary interaction strength, and the symmetry-preserving nature of the truncation. 

A possible way to 
improve on the present analysis is 
to extend the quark-gluon vertex 
in the complex plane, at least 
within a domain where it remains analytic, and features such as algebraic singularities or branch cuts have not made their appearance. Depending on the extension of this domain, one might 
be able to obtain some of the meson masses directly, without the need to 
resort to extrapolations. In fact, 
it is likely that 
even heavier masses may be attained by applying 
the SPM on data samples enriched with 
points in the Minkowski space. 
Indeed, even a limited amount of such points is bound to improve the implementation of 
the SPM, since the effective range of the extrapolation 
will be reduced, and 
the associated error will decrease.  Calculations in that direction are already in progress, and we hope to present results  
on this subject in the near future. 

\section*{Acknowledgments}\label{sec:Acknow}
A.S.M., J.M.M. and J.P. are funded by the Spanish MICINN grants PID2020-113334GB-I00 and PID2023-151418NB-I00, the Generalitat Valenciana grant CIPROM/2022/66, and CEX2023-001292-S by MCIU/AEI. Part of the calculations have been computed using the General Computing Infrastructure (GLUON) of the University of Valencia. We thank Markus Huber for sharing with us the data from \cite{Hagel:2025ngi}.

\appendix
\section{Renormalization with different current quark masses}\label{sec:ren}

The standard relations 
connecting bare and renormalized quantities 
are given by 
\begin{align}
S_{\!\s{R}f}(p) &= Z_{2f}^{-1} 
S_{f}(p) \,,&m_{\!\s{R}f} &= Z_{m_{f}}^{-1} m_{f} \,,&\nonumber\\
\Gamma_{\!\!{\s{R}}\,5\mu}^{ff'}(p_1, p_2) &= Z_{\!\s{A}}^{ff'} 
\Gamma_{5\mu}^{ff'}(p_1, p_2) \,,&
\Gamma_{\!\!{\s{R}}\, 5}^{ff'}(p_1, p_2) &= Z_{\!\s{P}}^{ff'}
\Gamma_{5}^{ff'}(p_1, p_2) \,.
\label{eq:ren}
\end{align}
The key requirement that the 
WTI maintains its form 
intact after renormalization imposes 
stringent relations among the 
various renormalization constants.
In particular, we have 
\be
Z_{2f} = Z_{2f'} = Z_{\!\s{A}}^{ff'} 
= Z_{m_{f}}^{-1} Z_{\!\s{P}}^{ff'} =
Z_{m_{f'}}^{-1} Z_{\!\s{P}}^{ff'} \,,\qquad 
Z_{m_{f}} = Z_{m_{f'}} \,.
\label{eq:constraints}
\ee
It is clear that the renormalization 
may not proceed through the standard 
MOM scheme~\cite{Celmaster:1979km,Hasenfratz:1980kn,Braaten:1981dv}. This is so because, in this scheme, 
the values for $Z_{2f}$ and $Z_{mf}$ are  
obtained from combinations of the 
integrals defining the 
quark self-energy $\Sigma_f(p)$, evaluated at 
the renormalization point $\mu$.
Since these integrals are flavour-dependent, the corresponding 
renormalization constants 
vary from one quark flavour to the next.

Instead, one must resort to renormalization schemes where the renormalization constants are flavour-independent, as required by \1eq{eq:constraints}. In perturbation theory, this is naturally accomplished through the MS schemes~\cite{Bollini:1972ui,Ashmore:1972uj,tHooft:1972tcz}, which, however, are not easily implemented nonperturbatively.

However, as first shown by Weinberg~\cite{Weinberg:1973xwm}, it is possible to modify the MOM schemes to render the renormalization constants 
flavour-independent; this approach is often employed in lattice studies~\cite{Martinelli:1994ty,Sturm:2009kb}.

The key observation is that the divergent parts of the renormalization constants are independent of the current masses~\cite{Weinberg:1973xwm}. Thus, we can impose for every quark flavour, $f$, that the corresponding renormalization constants $Z_{2f}$ and $Z_{m_{f}}$ have the same values as their counterparts for the lightest quark, $u$, \ie
\be 
Z_{2f} = Z_{2u} =: Z_2 \,, \qquad Z_{m_f} = Z_{m_u} =: Z_m \,. \label{Zs_eq}
\ee

With this prescription, the conditions in \1eq{eq:constraints} are exactly fulfilled. Note that the use of the quark $u$ as a reference to define all renormalization constants is arbitrary, and chosen to simplify the numerical calculations; any other quark mass, or even the chiral limit, would similarly define valid renormalization schemes~\cite{Weinberg:1973xwm,Pascual:1980yu,Martinelli:1994ty,Chetyrkin:1999pq,Sturm:2009kb}.

It remains to specify the values of $Z_{2}$ and $Z_{m}$. To this end, we begin by renormalizing the gap equation of \1eq{eq:gap} by means of the relations in \1eq{eq:ren}, to obtain
\be
S^{-1}(p)= Z_{2} \slashed{p} - Z_{2} Z_{m} m_f(\mu) - i\Sigma_f(p)\,,
\label{eq:gap_ren}
\ee
where $m_f$, just as the coupling, depends on the renormalization scale, $\mu$, and $\Sigma_f(p)$ now denotes the renormalized quark self-energy,
\begin{align}
\label{eq:Sigma_ren}
\Sigma_f(p)= - g^2C_fZ_{1f}\int_q\gamma^\nu S(q)\g^\mu(q-p,p,-q)\Delta_{\mu\nu}(q-p) \,,
\end{align}
with $Z_{1f}$ being the renormalization constant of the quark-gluon vertex. Note that, since $Z_{1f}$ does not appear in the constraints of \1eq{eq:constraints}, we can employ standard MOM conditions for the quark-gluon vertex (for details see, \eg \cite{Aguilar:2024ciu}). Then, taking traces, we find
\be
A_f(p) = Z_{2} - i\Sigma_{Af}(p)\,, \qquad B_f(p) = Z_{2}Z_{m} m_f(\mu) - i\Sigma_{Bf}(p)\,, \label{A_B_ren}
\ee
where we write $\Sigma_f(p) = \slashed{p} \Sigma_{Af}(p) - \Sigma_{Bf}(p)$, such that
\be
\Sigma_{Af}(p) = \frac{1}{4p^2}\Tr\left[ \slashed{p} \Sigma_f(p) \right]\,, \qquad \Sigma_{Bf}(p) = - \frac{1}{4}\Tr\left[ \Sigma_f(p) \right]\,.
\ee

At this point, we impose MOM conditions for the propagator, $S_u(p)$, of the quark $u$, at the renormalization point $p^2 = \mu^2$, \ie
\be 
A_u(\mu) = 1\,, \qquad B_u(\mu) = m_u(\mu) \,. \label{MOM_u}
\ee
Thus, setting $f = u$, and using \2eqs{MOM_u}{Zs_eq} in \1eq{A_B_ren}, fixes the values of $Z_2$ and $Z_m$, namely
\be
Z_{2} = 1 + i\Sigma_{Au}(\mu)\,, \qquad Z_{2}Z_{m} = Z_{2}Z_{m} = 1 + \frac{i\Sigma_{Bu}(\mu)}{m_u(\mu)}\,. \label{Zs_vals}
\ee

To complete, we substitute \1eq{Zs_vals} back in \1eq{A_B_ren} to obtain the subtracted expression
\be
A_f(p) = 1 - i\left[ \Sigma_{Af}(p) - \Sigma_{Au}(\mu) \right]\,, \qquad B_f(p) = m_f(\mu) - i\left[ \Sigma_{Bf}(p) - \frac{m_f(\mu)}{m_u(\mu)} \Sigma_{Bu}(\mu) \right] \,. \label{A_B_ren_f}
\ee
Note that for $f = u$, the above expressions reduce to the MOM conditions at $p^2 = \mu^2$. On the other hand, for $f \neq u$, $A_f(\mu)$ and $B_f(\mu)$ do not necessarily reduce to 1 and $m_f(\mu)$, respectively. 

As in any other renormalization scheme, the values of the renormalized coupling, $\alpha_s(\mu)$, and current masses, $m_u(\mu)$ and $m_s(\mu)$, are parameters of the theory, and are adjusted to reproduce observables. Specifically, we fix these parameters from the experimental values of the masses of the $\pi$ and $K$ mesons (see \tab{tab:meson_masses}), to find that $\alpha_s(\mu = 2\,\text{GeV}) = 0.513$, $m_u(\mu=2\,\text{GeV}) = 42$~MeV, and $m_s(\mu=2\,\text{GeV}) = 3.9$~GeV.

We hasten to emphasize that the $m_f(\mu)$ depend on the renormalization scheme and point adopted, and as such, have no direct physical meaning. In particular, in flavour-independent MOM schemes, such as the one described above, the renormalized current masses do not generally correspond to pole masses even in perturbation theory~\cite{Weinberg:1973xwm,Pascual:1980yu,Martinelli:1994ty,Chetyrkin:1999pq,Sturm:2009kb}. Thus, the large value of $m_s(\mu=2\,\text{GeV})$ obtained above 
has no bearing on the physics. 

Instead, a more meaningful quantity is the constituent mass, ${\cal M}_f$, which is RGI. Importantly, in spite of the large \mbox{$m_s(\mu=2\,\text{GeV}) = 3.9$~GeV}, we find that the constituent mass of the strange quark attains reasonable values, in particular, \mbox{${\cal M}_s(\mu=2\,\text{GeV}) = 155$~MeV}. Moreover, since ${\cal M}_s(\mu)$ is RGI, its value in the flavour independent scheme is the same as in the standard MOM scheme, \ie ${\cal M}_f(\mu) = {\cal M}_f^{\srm{MOM}}(\mu)$. Then, since by definition ${\cal M}_f^{\srm{MOM}}(\mu) = m_f^{\srm{MOM}}(\mu)$, we also have $m_f^{\srm{MOM}}(\mu) = 155$~MeV.

\begin{figure}[t]
    \hspace*{-1.5cm}
    \includegraphics[width=1.15\textwidth]{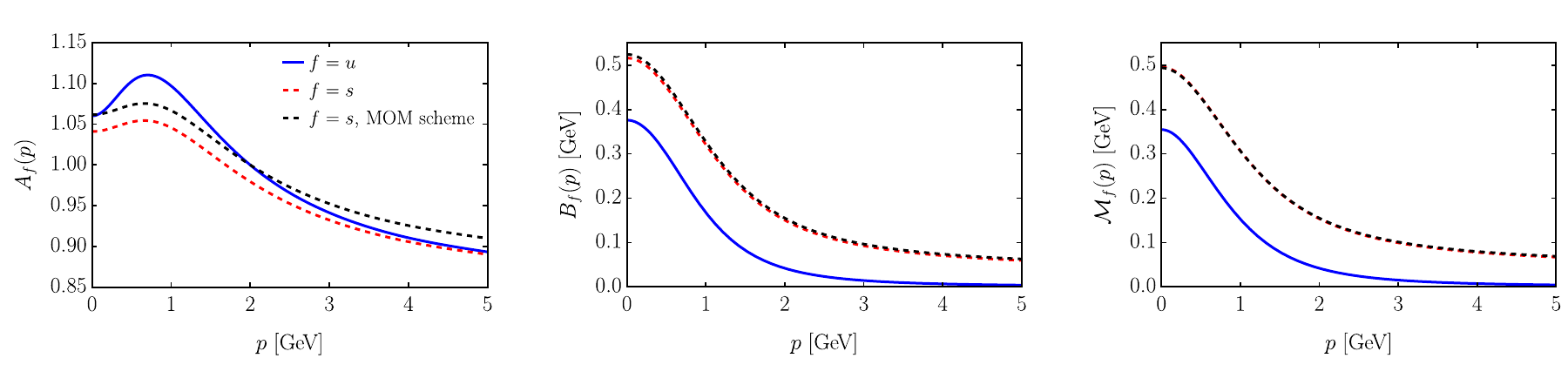}
    \caption{Solution of the gap equation for $A_f(p)$ (left), $B_f(p)$ (center), and the corresponding constituent mass, ${\cal M}_f(p)$ (right). The results in the flavour independent renormalization scheme for $f = u$ are shown as blue solid lines  and for $f = s$ as red dashed ones. The black dashed line represents the result for the strange quark in the standard MOM scheme.}
    \label{fig:quark_AB}
\end{figure}

In \fig{fig:quark_AB} we show the resulting $A_f(p)$ (left), $B_f(p)$ (center), and ${\cal M}_f(p)$ (right), for the $u$ and $s$ quark (blue continuous and red dashed lines, respectively). Note that, as expected, for $f = s$, $A_s(\mu) = 0.98$ and, especially, $B_s(\mu) = 150$~MeV do not satisfy MOM conditions. 

As a check on our renormalization procedures, we also show in \fig{fig:quark_AB} the results for the $s$ quark in the standard MOM scheme (black dashed). The latter is obtained by solving the gap equation with MOM conditions also for $s$, \ie
\be 
A_s^{\srm{MOM}}(\mu) = 1\,, \qquad B_s^{\srm{MOM}}(\mu) = m_s^{\srm{MOM}}(\mu) \,, \label{MOM_s}
\ee
which entails
\be
A_s^{\srm{MOM}}(p) = 1 - i\left[ \Sigma_{As}^{\srm{MOM}}(p) - \Sigma_{As}^{\srm{MOM}}(\mu) \right]\,, \qquad B_s^{\srm{MOM}}(p) = m_s^{\srm{MOM}}(\mu) - i\left[ \Sigma_{Bs}^{\srm{MOM}}(p) - \Sigma_{Bs}^{\srm{MOM}}(\mu) \right] \,, \label{A_B_ren_s}
\ee
where we use the value $m_s^{\srm{MOM}}(\mu=2\,\text{GeV}) = 155$~MeV, determined above. Importantly, while $A_s(p)$ and $B_s(p)$ are different from their MOM values, ${\cal M}_s(p) = {\cal M}_s^{\srm{MOM}}(p)$, as required.

\clearpage
\bibliography{bibliography.bib}

\end{document}